\documentclass[a4paper,12pt]{article}
\usepackage{amssymb}

\begin{document}

\title{Uniqueness of the Machian Solution in the Brans-Dicke Theory}
\author{A. Miyazaki \thanks{
Email: miyazaki@loyno.edu, miyazaki@nagasakipu.ac.jp} \vspace{3mm} \\
\textit{Department of Physics, Loyola University, New Orleans, LA 70118} \\
and \\
\textit{Faculty of Economics, Nagasaki Prefectural University} \\
\textit{Sasebo, Nagasaki 858-8580, Japan}}
\date{\vfill}
\maketitle

\begin{abstract}
Machian solutions of which the scalar field exhibits the asymptotic behavior 
$\phi =O(\rho /\omega )$ are generally explored for the homogeneous and
isotropic universe in the Brans-Dicke theory. It is shown that the Machian
solution is unique for the closed and the open space. Such a solution is
restricted to one that satisfies the relation $GM/c^{2}a=const$, which is
fixed to $\pi $ in the theory for the closed model. Another type of solution
satisfying $\phi =O(\rho /\omega )$ with the arbitrary coupling constant $%
\omega $ is obtained for the flat space. This solution has the scalar field $%
\phi \propto \rho \,t^{2}$ and also keeps the relation $GM/c^{2}R=const$ all
the time. This Machian relation and the asymptotic behavior $\phi =O(\rho
/\omega )$ is equivalent to each other in the Brans-Dicke theory.\newline
\end{abstract}

\newpage

\section{Introduction}

In the previous paper \cite{1)}, we discussed the asymptotic behavior of the
scalar field $\phi $\ of the Brans-Dicke theory \cite{2)} for the large
enough coupling parameter $\omega $, and proposed two postulates in the
Machian point of view: The scalar field of a proper cosmological solution
should have the asymptotic form $\phi =O(\rho /\omega )$ and should converge
to zero in the continuous limit $\rho /\omega \rightarrow 0$. (Let us call
it ''the Machian solution'' here.) The scalar field by locally-distributed
matter should exhibit the asymptotic behavior $\phi =\left\langle \phi
\right\rangle +O(1/\omega )$. When we consider local problems in the
Brans-Dicke theory, we always premise tacitly\ the presence of cosmological
matter in the universe. If we restrict to the scalar field with the
asymptotic behavior $\phi =\left\langle \phi \right\rangle +O(1/\omega )$,
we can discuss possible solutions for locally-distributed matter with an
asymptotically-flat boundary condition for the metric tensor $g_{\mu \nu }$
and with another boundary condition that the local scalar field $\phi $
converges to $\left\langle \phi \right\rangle $, which is given by the
experimental gravitational constant $G$, at the asymptotic region ($%
r\rightarrow +\infty $). We may forget our cosmological environment on this
assumption.

Taking the large coupling parameter $\omega $\ restricted by the recent
measurements \cite{3)} ($\left| \omega \right| \gtrsim 10^{3}$) into
account, there exists only an extremely small difference between general
relativity and the Brans-Dicke theory for local problems.\ However, there
appears a crucial difference for cosmological problems, especially in the
Machian point of view. It becomes more important to survey systematically
such Machian solutions in the Brans-Dicke theory. What is a cosmological
solution with the asymptotic form $\phi =O(\rho /\omega )$ in the
Brans-Dicke theory? The purpose of this article is seeking the general
existence of solutions satisfying asymptotically $\phi =O(\rho /\omega )$\
in the Brans-Dicke theory.

Such an example is already known \cite{4)} and \cite{5)}, \cite{6)} (by
satisfying $a\phi =const$) for the closed model, and \cite{7)} for the flat
and the open models. For the homogeneous and isotropic universe with the
perfect-fluid matter (with negligible pressure), 
\begin{equation}
ds^{2}=-dt^{2}+a^{2}(t)[d\chi ^{2}+\sigma ^{2}(\chi )(d\theta ^{2}+\sin
^{2}\theta d\varphi ^{2})]\,,  \label{e1}
\end{equation}
where 
\begin{equation}
\sigma (\chi )\equiv \left\{ 
\begin{array}{l}
\sin \chi \;\;\;\;\;for\;k=+1\;(closed\;space) \\ 
\chi \;\;\;\;\;\;\;\;for\;k=0\;\ \;(flat\;space) \\ 
\sinh \chi \;\;\;for\;k=-1\;\ (open\;space)\,,
\end{array}
\right.  \label{e2}
\end{equation}
the field equations of the Brans-Dicke theory has the following cosmological
solutions: 
\begin{equation}
a(t)=\left[ -4M/(3+2\omega )\pi c^{2}D\right] ^{1/2}t\,,  \label{e3}
\end{equation}
\begin{equation}
2\pi ^{2}a^{3}(t)\rho (t)=M\,,  \label{e4}
\end{equation}
\begin{equation}
a(t)\phi (t)=D  \label{e5}
\end{equation}
on the condition that 
\begin{equation}
\omega /(3+2\omega )=2-k(3\pi c^{2}D/2M)  \label{e6}
\end{equation}
with 
\begin{equation}
\begin{array}{l}
-2<\omega \;\;(\omega \neq -3/2)\;\;\;for\;k=-1, \\ 
\omega =-2\,,\;D>0\;\;\;\;\;\;\;\;for\;k=0\,, \\ 
\omega <-2\,,\;M/c^{2}D>\pi \;\;for\;k=+1\,,
\end{array}
\label{e7}
\end{equation}
where $D$ and $M$ are constants ($M$ means the total mass of the universe
for the closed model),\ and $\rho $ is the mass density.

For the open model, the expansion parameter Eq.(\ref{e3})\ is rewritten
simply to 
\begin{equation}
a(t)=[2/(2+\omega )]^{1/2}t\,,  \label{e8}
\end{equation}
and for the closed model 
\begin{equation}
a(t)=[-2/(2+\omega )]^{1/2}t\,.  \label{e9}
\end{equation}
It should be noted that only the coupling parameter $\omega $ determines the
behavior of the expansion parameter. For the flat model, the coefficient of
the expansion parameter $a(t)$\ becomes indefinite. The scalar field $\phi $%
\ has explicitly the following form for all cases ($k=0,\,\pm 1$): 
\begin{equation}
\phi (t)=-[8\pi /(3+2\omega )c^{2}]\rho (t)t^{2}\,.  \label{e10}
\end{equation}
Taking the region of $\omega $ in Eq.(\ref{e7}) and the relation between the
gravitational constant $G$ and the scalar field $\phi $, 
\begin{equation}
G\equiv (4+2\omega )/(3+2\omega )\phi \,,  \label{e11}
\end{equation}
into account, we find signs of the gravitational constant in these solutions 
\begin{equation}
G<0\,,G=0\,,G>0\,,  \label{e12}
\end{equation}
for $k=-1$, $k=0$, and $k=+1$ respectively.

The closed cosmological solution satisfying the additional condition $a\phi
=const$, which is equivalent to $GM/c^{2}a=const$,\ exists only for its
value 
\begin{equation}
\frac{G(t)M}{c^{2}a(t)}=\pi \,.  \label{e13}
\end{equation}
Therefore, the well-known Machian relation 
\begin{equation}
\frac{GM}{c^{2}R}\sim 1\,,  \label{e14}
\end{equation}
where $R$\ is the radius of the universe, is automatically satisfied in this
cosmological model all the time. However, this relation is ambiguous as to
the meaning of the radius $R$\ or the mass $M$\ of the universe, and might
not be essential to discuss Machian cosmological models.

\ According to the previous paper \cite{1)}, let us adopt the following
relation of the scalar field 
\begin{equation}
\phi =O(\rho /\omega )\,  \label{e15}
\end{equation}
as the more fundamental Machian relation in the Brans-Dicke theory. When the
mass density $\rho $\ of the universe goes to zero or the coupling between
the scalar field and matter vanishes ($\omega \rightarrow \infty $), the
scalar field $\phi $\ also converges to zero. This means that a particle in
any other empty space does not have the inertial properties. Thus the scalar
field $\phi $\ should be described as 
\begin{equation}
\phi (t)=\frac{8\pi \rho (t)}{(3+2\omega )c^{2}}\Psi (t)\,,  \label{e16}
\end{equation}
where $\Psi $ is a function of $t$\ and may depend on $\omega $\ as 
\begin{equation}
\Psi (t)=\Psi _{0}(t)+O(1/\omega )\,.  \label{e17}
\end{equation}
We, however, find that it is difficult and inappropriate to seek this type
of solutions. Let us start simply from another relation 
\begin{equation}
\phi (t)=\frac{8\pi }{(3+2\omega )c^{2}}\Phi (t)\,.  \label{e18}
\end{equation}
A function $\Phi $ may also have the same dependence of $\omega $\ as Eq.(%
\ref{e17}).

\section{Basic Equations and the Flat-Space Case}

For the homogeneous and isotropic metric Eq.(\ref{e1}), the nonvanishing
component of the energy-momentum tensor of the perfect-fluid with negligible
pressure is $T_{00}=-\rho c^{2}$, and the contracted energy-momentum tensor
is $T=\rho c^{2}$.\ Thus the nonvanishing field equations we need solve
simultaneously in the Brans-Dicke theory are 
\begin{equation}
2\dot{a}\ddot{a}+\dot{a}^{2}+k=-\frac{8\pi }{(3+2\omega )c^{2}}\frac{%
a^{2}\rho }{\phi }-\frac{1}{2}\omega a^{2}\left( \frac{\dot{\phi}}{\phi }%
\right) ^{2}+a\dot{a}\left( \frac{\dot{\phi}}{\phi }\right) \,,  \label{e19}
\end{equation}
\begin{equation}
\frac{3}{a^{2}}\left( \dot{a}^{2}+k\right) =\frac{16\pi (1+\omega )}{%
(3+2\omega )c^{2}}\frac{\rho }{\phi }+\frac{\omega }{2}\left( \frac{\dot{\phi%
}}{\phi }\right) ^{2}+\frac{\ddot{\phi}}{\phi }\,,  \label{e20}
\end{equation}
\begin{equation}
\ddot{\phi}+3\frac{\dot{a}}{a}\dot{\phi}=\frac{8\pi }{(3+2\omega )c^{2}}\rho
\,,  \label{e21}
\end{equation}
where a dot denotes the partial derivative with respect to $t$.\ However, we
can use the conservation law of the energy-momentum, which is derived from
the field equations, as the independent equation instead of Eq.(\ref{e19}),
and obtain for this case 
\begin{equation}
2\pi ^{2}a^{3}(t)\rho (t)=M\,.  \label{e22}
\end{equation}

If we substitute Eq.(\ref{e18}) into Eq.(\ref{e21}), we get directly 
\begin{equation}
\ddot{\Phi}+3\frac{\dot{a}}{a}\dot{\Phi}=\rho \,.  \label{e23}
\end{equation}
Let us suppose the mass density $\rho $\ and the function $\Phi $\ do not
depend on the coupling parameter $\omega $, then we observe that the ratio $%
\dot{a}/a$ also does not depend $\omega $:\ 
\begin{equation}
\frac{\dot{a}}{a}\equiv \gamma (t)\,,  \label{e24}
\end{equation}
where a function $\gamma $\ depends on only $t$.\ By integrating Eq.(\ref
{e24}), we find the expansion parameter should have a form 
\begin{equation}
a(t)\equiv A(\omega )\alpha (t)\,,  \label{e25}
\end{equation}
where $A$ and $\alpha $\ are arbitrary functions of only $\omega $\ and $t$\
respectively.

From Eqs.(\ref{e21}) and (\ref{e22}) after integration, we obtain 
\begin{equation}
\dot{\phi}a^{3}=[4M/(3+2\omega )\pi c^{2}]t\,,  \label{e26}
\end{equation}
where we take a constant of integration into zero. From Eqs.(\ref{e20}), (%
\ref{e21}), (\ref{e22}), and (\ref{e26}), we get\ 
\begin{equation}
\left[ \left( \frac{\dot{a}}{a}\right) +\frac{1}{2}\left( \frac{\dot{\phi}}{%
\phi }\right) \right] ^{2}+\frac{k}{a^{2}}=\frac{1}{4}\left( 1+\frac{2\omega 
}{3}\right) \left[ \left( \frac{\dot{\phi}}{\phi }\right) ^{2}+4\left( \frac{%
\dot{\phi}}{\phi }\right) \frac{1}{t}\right] \,.  \label{e27}
\end{equation}
Similarly we obtain 
\begin{equation}
\dot{\Phi}a^{3}=(M/2\pi ^{2})t\,  \label{e28}
\end{equation}
and 
\begin{equation}
\left[ \left( \frac{\dot{a}}{a}\right) +\frac{1}{2}\left( \frac{\dot{\Phi}}{%
\Phi }\right) \right] ^{2}+\frac{k}{a^{2}}=\frac{1}{4}\left( 1+\frac{2\omega 
}{3}\right) \left[ \left( \frac{\dot{\Phi}}{\Phi }\right) ^{2}+4\left( \frac{%
\dot{\Phi}}{\Phi }\right) \frac{1}{t}\right] \,.  \label{e29}
\end{equation}
It should be noted that the scalar field $\phi $ appears only as the ratio $%
\dot{\phi}/\phi $ in Eq.(\ref{e27}).

We know the general exact solution of Eqs.(\ref{e26}) and (\ref{e27}) for
the flat-space case \cite{2)}, and this solution is also valid for the early
expansion phases for the closed or the open spaces. Therefore, Eqs.(\ref{e28}%
) and (\ref{e29}) has the similar general solution in the same situation: 
\begin{equation}
\Phi =\Phi _{0}(t/t_{0})^{r}\,,\;\;r=2/(4+3\omega )\,,  \label{e30}
\end{equation}
\begin{equation}
\Phi _{0}=[(4+3\omega )/2]\rho _{0}t_{0}^{2}\,.  \label{e31}
\end{equation}
In the result, Eq.(\ref{e18}) has become the usual non-Machian cosmological
solution of which the asymptotic behavior is 
\begin{equation}
\phi =\left\langle \phi \right\rangle +O(1/\omega )\,.  \label{e32}
\end{equation}
Thus we conclude that the function $\Phi $ should not depend on the coupling
parameter $\omega $ for the Machian cosmological solution. Equation (\ref
{e18}) with the function $\Phi (t)$ depending on only $t$\ is the most
general form for the Machian solution $\phi =O(\rho /\omega )$. The mass
density $\rho (t)$\ is given first and should not also depend on $\omega $.
So we confirm that the expansion parameter $a(t)$ is described like Eq.(\ref
{e25}). The $\omega $-dependence of the expansion parameter Eq.(\ref{e25})
is put into the total mass of the universe $M(\omega )$. Finally we obtain
from Eqs.(\ref{e20}), (\ref{e21}), (\ref{e18}), and (\ref{e25}) 
\begin{equation}
\frac{\omega }{2}\left[ \left( \frac{\dot{\Phi}}{\Phi }\right) ^{2}+\frac{%
4\rho }{\Phi }\right] -\frac{3k}{A^{2}(\omega )\alpha ^{2}}=3\left( \frac{%
\dot{\alpha}}{\alpha }\right) ^{2}+3\left( \frac{\dot{\alpha}}{\alpha }%
\right) \left( \frac{\dot{\Phi}}{\Phi }\right) -\frac{3\rho }{\Phi }\,.
\label{e33}
\end{equation}
The independent equations which we solve simultaneously are Eqs.(\ref{e33}),
(\ref{e28}), and (\ref{e22}) for $k=0,\,\pm 1$.

Let us discuss the flat model ($k=0$) first. If we request that Eq.(\ref{e33}%
) is always satisfied for all arbitrary values of $\omega $, we find the two
following conditions must be held identically,\ 
\begin{equation}
\left( \frac{\dot{\Phi}}{\Phi }\right) ^{2}+\frac{4\rho }{\Phi }\equiv 0\,,
\label{e34}
\end{equation}
and 
\begin{equation}
\left( \frac{\dot{\alpha}}{\alpha }\right) ^{2}+\left( \frac{\dot{\alpha}}{%
\alpha }\right) \left( \frac{\dot{\Phi}}{\Phi }\right) -\frac{\rho }{\Phi }%
\equiv 0\,.  \label{e35}
\end{equation}
After eliminating the term $\rho /\Phi $ from Eqs.(\ref{e34}) and (\ref{e35}%
), we can resolve the result into factors and obtain 
\begin{equation}
\left[ \left( \frac{\dot{\Phi}}{\Phi }\right) +2\left( \frac{\dot{\alpha}}{%
\alpha }\right) \right] ^{2}=0\,.  \label{e36}
\end{equation}
The integral of the term in the middle bracket gives 
\begin{equation}
\Phi (t)\alpha ^{2}(t)=const\,.  \label{e37}
\end{equation}

On the other hand, we get from Eqs.(\ref{e34}), (\ref{e28}), and (\ref{e22}) 
\begin{equation}
\Phi (t)=-\frac{M}{8\pi ^{2}A^{3}}\frac{t^{2}}{\alpha ^{3}(t)}\,.
\label{e38}
\end{equation}
If we eliminate the term $\alpha ^{3}$ from Eq.(\ref{e38}) using Eq.(\ref
{e28}), we obtain simply 
\begin{equation}
\frac{\dot{\Phi}}{\Phi }=-\frac{4}{t}\,.  \label{e39}
\end{equation}
We can integrate this equation exactly and find 
\begin{equation}
\Phi (t)\propto t^{-4}\,,  \label{e40}
\end{equation}
Substituting this equation into Eq.(\ref{e38}) yields 
\begin{equation}
\alpha (t)\propto t^{2}\,.  \label{e41}
\end{equation}
Though time-dependences of $\Phi (t)$ and $\alpha (t)$ are compatible with
Eq.(\ref{e37}), we can not determine the coefficient of $\alpha (t)$.
Equations (\ref{e34}), (\ref{e35}), and (\ref{e23}) are no more independent
to each other and the solution becomes indefinite. It is also obvious that
the coefficient of the expansion parameter $a(t)$ becomes indefinite because
only the ratio $\dot{a}/a$ appears in Eqs.(\ref{e23}) and (\ref{e29}) for
the flat-space case ($k=0$). The coefficient of $a(t)$ is also indefinite in
the particular solution (\ref{e3}) with $\omega =-2$ for the flat space. It
seems that this indefiniteness is the situation characteristic of the
flat-space case.

If we fix a value $a_{0}$ of the expansion parameter at an arbitrary time $%
t_{0}$, we may write down the indefinite solution as 
\begin{equation}
a(t)=a_{0}\left( t/t_{0}\right) ^{2}  \label{eq1}
\end{equation}
and the scalar field $\phi (t)$ is given by the mass density $\rho (t)$ as 
\begin{equation}
\phi (t)=-[2\pi /(3+2\omega )c^{2}]\rho (t)t^{2}\,,  \label{eq2}
\end{equation}
which is valid for all arbitrary values ($\omega \neq -3/2$) of the coupling
parameter $\omega $ ($\omega <-2$ for $G>0$). If we restrict to the
causally-related region, we observe the following relation \emph{for all} $t$%
, taking the radius of the horizon $R(t)=t$ and the total mass inside the
horizon $M(t)=(4/3)\pi R^{3}(t)\rho (t)$ into account: 
\begin{equation}
\frac{G(t)M(t)}{c^{2}R(t)}=-4(2+\omega )/3=const\,.  \label{eq3}
\end{equation}
It is essential for the relation $GM/c^{2}R=const$ that the scalar field $%
\phi $ has the form $\rho (t)t^{2}$, which is derived from Eq.(\ref{e34}).

\section{Closed-Space Case}

For the closed or the open models ($k=\pm 1$), we obtain from Eq.(\ref{e33})
the two constraints: 
\begin{equation}
\frac{\omega }{2}\left[ \left( \frac{\dot{\Phi}}{\Phi }\right) ^{2}+\frac{%
4\rho }{\Phi }\right] -\frac{3k}{A^{2}(\omega )\alpha ^{2}}\equiv C(t)\,,
\label{e42}
\end{equation}
\begin{equation}
3\left( \frac{\dot{\alpha}}{\alpha }\right) ^{2}+3\left( \frac{\dot{\alpha}}{%
\alpha }\right) \left( \frac{\dot{\Phi}}{\Phi }\right) -\frac{3\rho }{\Phi }%
\equiv C(t)\,,  \label{e43}
\end{equation}
where $C$ is a function of only $t$. In order that Eq.(\ref{e42}) is
satisfied for all arbitrary values of $\omega $, the function $A$ must have
the following form\ 
\begin{equation}
\frac{3}{A^{2}(\omega )}=\left| \frac{\omega }{2}+B\right| \,,  \label{e44}
\end{equation}
where $B$\ is a constant with no dependence of $\omega $. Therefore, in the
Machian solutions for the homogeneous and isotropic universe (except the
flat space), the expansion parameter $a(t)$ exhibits generally the
asymptotic behavior $O(1/\sqrt{\omega })$ for the large enough coupling
parameter $\omega $.

For $k=+1,$ and $\omega /2+B<0$, Eq.(\ref{e44}) gives 
\begin{equation}
\frac{3}{A^{2}(\omega )}=-\left( \frac{\omega }{2}+B\right) \,,  \label{e45}
\end{equation}
and we get from Eq.(\ref{e33}) 
\begin{equation}
\frac{\omega }{2}\left[ \left( \frac{\dot{\Phi}}{\Phi }\right) ^{2}+\frac{%
4\rho }{\Phi }+\frac{1}{\alpha ^{2}}\right] =3\left( \frac{\dot{\alpha}}{%
\alpha }\right) ^{2}+3\left( \frac{\dot{\alpha}}{\alpha }\right) \left( 
\frac{\dot{\Phi}}{\Phi }\right) -\frac{3\rho }{\Phi }-\frac{B}{\alpha ^{2}}%
\,.  \label{e46}
\end{equation}
We need to hold the two following identities to satisfy this equation for
all arbitrary $\omega $: 
\begin{equation}
\left( \frac{\dot{\Phi}}{\Phi }\right) ^{2}+\frac{4\rho }{\Phi }\equiv -%
\frac{1}{\alpha ^{2}}\,,  \label{e47}
\end{equation}
\begin{equation}
3\left( \frac{\dot{\alpha}}{\alpha }\right) ^{2}+3\left( \frac{\dot{\alpha}}{%
\alpha }\right) \left( \frac{\dot{\Phi}}{\Phi }\right) -\frac{3\rho }{\Phi }%
\equiv \frac{B}{\alpha ^{2}}\,.  \label{e48}
\end{equation}

If we eliminate the derivative $\dot{\Phi}$ from Eq.(\ref{e47}) using Eq.(%
\ref{e28}), we obtain 
\begin{equation}
\Phi ^{2}(t)+\frac{2M}{\pi ^{2}A^{3}\alpha (t)}\Phi (t)+\left( \frac{Mt}{%
2\pi ^{2}A^{3}\alpha ^{2}(t)}\right) ^{2}=0\,,  \label{e49}
\end{equation}
and find directly 
\begin{equation}
\Phi (t)=-\frac{M}{\pi ^{2}A^{3}\alpha (t)}\left( 1\pm \sqrt{1-t^{2}/4\alpha
^{2}(t)}\right) \,.  \label{e50}
\end{equation}
In order that the function $\Phi $\ is real for all times ($0\leqq t<+\infty 
$), the time-dependence of the expansion parameter must be at least linear: 
\begin{equation}
\alpha (t)\propto t\,.  \label{e51}
\end{equation}
Therefore, we observe from Eq.(\ref{e50}) 
\begin{equation}
\Phi (t)\alpha (t)=const\,,  \label{e52}
\end{equation}
and hence 
\begin{equation}
\Phi (t)\propto t^{-1}\,.  \label{e53}
\end{equation}
We can determine the expansion parameter $a^{2}(t)=-[2/(\omega +2)]t$
directly from Eq.(\ref{e29}) and then find easily the coefficient of $\alpha
(t)$ and the value of $B$\ satisfying Eq.(\ref{e48}): 
\begin{equation}
\alpha (t)=t\,/\sqrt{3}\,,\;\;B=1\,,  \label{54}
\end{equation}
and we get 
\begin{equation}
\Phi (t)\alpha (t)=-\frac{3M}{2\pi ^{2}A^{3}}\,,  \label{e55}
\end{equation}
which means 
\begin{equation}
\Phi (t)=-\rho (t)t^{2}\,.  \label{e56}
\end{equation}
This is nothing but the Machian solution Eq.(\ref{e10}) for the closed space
($k=+1$).

Next we discuss the possibility of the bounded universe with a finite time.
We restrict our discussions to the early expansion phases, and let us write 
\begin{equation}
\alpha (t)=bt^{\,\beta }\,,  \label{e57}
\end{equation}
where $b$ and $\beta $ are constants. When $\beta >1$, the second term of
Eq.(\ref{e50}) 
\begin{equation}
\sqrt{1-1/4b^{2}t^{2(\beta -1)}}  \label{e58}
\end{equation}
diverges at $t=0$, and there exist no solutions in this neighborhood. When$\
\beta <1$, for early expansion phases, the relation 
\begin{equation}
\alpha (t)\gg t\,  \label{e59}
\end{equation}
is valid, and we obtain from Eq.(\ref{e50}) 
\begin{equation}
\Phi (t)=-\frac{2M}{\pi ^{2}A^{3}\alpha (t)}\,,\;\;or\;\;0\,,  \label{e60}
\end{equation}
which does not satisfy Eq.(\ref{e48}). Thus, when $\beta \neq 1$, the
Machian solution does not exist for a finite and meaningful time in this
case ($k=+1$, $\omega /2+B<0$).

For $k=+1$, and $\omega /2+B>0$, the coefficient $A(\omega )$ has the form 
\begin{equation}
\frac{3}{A^{2}(\omega )}=\frac{\omega }{2}+B\,,  \label{e61}
\end{equation}
and Eq.(\ref{e33}) yields 
\begin{equation}
\frac{\omega }{2}\left[ \left( \frac{\dot{\Phi}}{\Phi }\right) ^{2}+\frac{%
4\rho }{\Phi }-\frac{1}{\alpha ^{2}}\right] =3\left( \frac{\dot{\alpha}}{%
\alpha }\right) ^{2}+3\left( \frac{\dot{\alpha}}{\alpha }\right) \left( 
\frac{\dot{\Phi}}{\Phi }\right) -\frac{3\rho }{\Phi }+\frac{B}{\alpha ^{2}}%
\,,  \label{e62}
\end{equation}
which produces, through the similar discussions for all arbitrary $\omega ,$%
\ the two following identities: 
\begin{equation}
\left( \frac{\dot{\Phi}}{\Phi }\right) ^{2}+\frac{4\rho }{\Phi }\equiv \frac{%
1}{\alpha ^{2}}\,,  \label{e63}
\end{equation}
\begin{equation}
3\left( \frac{\dot{\alpha}}{\alpha }\right) ^{2}+3\left( \frac{\dot{\alpha}}{%
\alpha }\right) \left( \frac{\dot{\Phi}}{\Phi }\right) -\frac{3\rho }{\Phi }%
\equiv -\frac{B}{\alpha ^{2}}\,.  \label{e64}
\end{equation}

The elimination of the derivative $\dot{\Phi}$ from Eq.(\ref{e63}) by Eq.(%
\ref{e28}) yields 
\begin{equation}
\Phi ^{2}(t)-\frac{2M}{\pi ^{2}A^{3}\alpha (t)}\Phi (t)-\left( \frac{Mt}{%
2\pi ^{2}A^{3}\alpha ^{2}(t)}\right) ^{2}=0\,,  \label{e65}
\end{equation}
which is solved directly to 
\begin{equation}
\Phi (t)=\frac{M}{\pi ^{2}A^{3}\alpha (t)}\left( 1\pm \sqrt{1+t^{2}/4\alpha
^{2}(t)}\right) \,.  \label{e66}
\end{equation}
It is enough to indicate that no solutions exist in the neighborhood of $t=0$
in order to prove that the Machian solution does not exist in this case, We
may restrict our discussions to the early expansion phases, and suppose Eq.(%
\ref{e57}).

When $\beta >1$ and $\alpha (t)\ll t$, we get from Eq.(\ref{e66}) 
\begin{equation}
\Phi (t)=\pm \frac{Mt}{2\pi ^{2}A^{3}\alpha ^{2}(t)}\,,  \label{e67}
\end{equation}
and find from Eqs.(\ref{e28}) and (\ref{e67}) after integration 
\begin{equation}
\Phi (t)=\Phi _{0}\exp \left[ \pm 1/(1-\beta )bt^{\beta -1}\right] \,,
\label{e68}
\end{equation}
which does not obviously satisfy Eq.(\ref{e64}).

When $\beta <1$ and $\alpha (t)\gg t$, from Eq.(\ref{e66}) 
\begin{equation}
\Phi (t)=\frac{2M}{\pi ^{2}A^{3}\alpha (t)}\,,\;\;or\;\;0\,.  \label{e69}
\end{equation}
For this approximation ($\alpha (t)\gg t$), we obtain Eqs.(\ref{e34}) and (%
\ref{e35}) from Eqs.(\ref{e63}) and (\ref{e64}) respectively and find the
constraint Eq.(\ref{e37}), which is not compatible with Eq.(\ref{e69}).

When $\beta =1$, Eq.(\ref{e66}) gives 
\begin{equation}
\Phi (t)=\frac{M}{\pi ^{2}A^{3}\alpha (t)}\left( 1\pm \sqrt{1+1/4b^{2}}%
\right) \,.  \label{e70}
\end{equation}
We can indicate straightforwardly that this equation does not satisfy Eqs.(%
\ref{e64}) and (\ref{e63}) or (\ref{e28}). Actually, if we put $b=1/\sqrt{3}$%
, then we obtain\newline
\begin{equation}
\Phi (t)=-\frac{M}{\pi ^{2}A^{3}\alpha (t)}\left( \sqrt{7}/2-1\right) <0\,,
\label{e71}
\end{equation}
which is not consistent with Eq.(\ref{e55}).

\section{Open-Space Case}

In similar fashion, the coefficient $A(\omega )$ of the expansion parameter $%
a(t)$ need to have the following form for the open model ($k=-1$) and $%
\omega /2+B>0$: 
\begin{equation}
\frac{3}{A^{2}(\omega )}=\frac{\omega }{2}+B\,.  \label{e72}
\end{equation}
We obtain from Eq.(\ref{e33}) taking this equation into account 
\begin{equation}
\frac{\omega }{2}\left[ \left( \frac{\dot{\Phi}}{\Phi }\right) ^{2}+\frac{%
4\rho }{\Phi }+\frac{1}{\alpha ^{2}}\right] =3\left( \frac{\dot{\alpha}}{%
\alpha }\right) ^{2}+3\left( \frac{\dot{\alpha}}{\alpha }\right) \left( 
\frac{\dot{\Phi}}{\Phi }\right) -\frac{3\rho }{\Phi }-\frac{B}{\alpha ^{2}}%
\,,  \label{e73}
\end{equation}
which is completely the same equation as in the case of $k=+1$, and $\omega
/2+B<0$. Again, therefore, we observe as the solution 
\begin{equation}
\alpha (t)=t\,/\sqrt{3}\,,\;\;B=1\,,  \label{e74}
\end{equation}
\begin{equation}
\Phi (t)\alpha (t)=-\frac{3M}{2\pi ^{2}A^{3}}\,,  \label{e75}
\end{equation}
and 
\begin{equation}
\Phi (t)=-\rho (t)t^{2}\,.  \label{e76}
\end{equation}

For $k=-1$, and $\omega /2+B<0$, the coefficient $A(\omega )$ has the form 
\begin{equation}
\frac{3}{A^{2}(\omega )}=-\left( \frac{\omega }{2}+B\right) \,,  \label{e77}
\end{equation}
and we get from Eq.(\ref{e33}) 
\begin{equation}
\frac{\omega }{2}\left[ \left( \frac{\dot{\Phi}}{\Phi }\right) ^{2}+\frac{%
4\rho }{\Phi }-\frac{1}{\alpha ^{2}}\right] =3\left( \frac{\dot{\alpha}}{%
\alpha }\right) ^{2}+3\left( \frac{\dot{\alpha}}{\alpha }\right) \left( 
\frac{\dot{\Phi}}{\Phi }\right) -\frac{3\rho }{\Phi }+\frac{B}{\alpha ^{2}}%
\,,  \label{e78}
\end{equation}
which is also completely the same as the case of $k=+1$, and $\omega /2+B>0$%
. We have already known that the Machian solution does not exist in this
region.

Finally, we discuss the case that the coefficient $A(\omega )$ of the
expansion parameter $a(t)$ does not depend on the coupling parameter $\omega 
$ for the closed and the open spaces ($k=\pm 1$). Let us suppose that the
function $\alpha (t)$\ absorbs the constant $A$\ and then we obtain from Eq.(%
\ref{e33}) 
\begin{equation}
\left( \frac{\dot{\Phi}}{\Phi }\right) ^{2}+\frac{4\rho }{\Phi }\equiv 0\,,
\label{e79}
\end{equation}
and 
\begin{equation}
\left( \frac{\dot{\alpha}}{\alpha }\right) ^{2}+\left( \frac{\dot{\alpha}}{%
\alpha }\right) \left( \frac{\dot{\Phi}}{\Phi }\right) -\frac{\rho }{\Phi }%
\equiv -\frac{k}{\alpha ^{2}}\,.  \label{e80}
\end{equation}
From Eqs.(\ref{e79}), (\ref{e28}), and (\ref{e22}), we find after
integration in the same way as the flat-space case 
\begin{equation}
\Phi (t)\propto t^{-4}\,,  \label{e81}
\end{equation}
and hence 
\begin{equation}
\alpha (t)\propto t^{2}\,.  \label{e82}
\end{equation}
Equations (\ref{e81}) and (\ref{e82}) does not satisfy Eq.(\ref{e80}) for $%
k=\pm 1$. Therefore, we observe that such a Machian solution does not exist
in the Brans-Dicke theory.

\section{Concluding Remarks}

We, after all, observe that the Machian solutions satisfying Eq.(\ref{e18})
have always the form $\Phi (t)\propto \rho (t)t^{2}$. This means that the
relation $GM/c^{2}a=const$ (or $GM/c^{2}R=const$ for the flat space) are
held \emph{for all} $t$ in these cosmological models for the homogeneous and
isotropic universe in the Brans-Dicke theory. They are equivalent to each
other, the relation $GM/c^{2}R=const$ (Let us include $GM/c^{2}a=const$.)
and the postulate that the scalar field exhibits the asymptotic behavior $%
\phi =O(\rho /\omega )$. It should be noted that the function $\Phi (t)$
does not depend on the coupling parameter $\omega $ in the Machian solution.
The solution (\ref{e10}), or Eq.(\ref{eq2}), has only the lowest order of $%
\rho /(3+2\omega )$ and is exact for all values of $\omega $.

As for the closed and the open spaces, the Machian solution with the scalar
field $\phi =O(\rho /\omega )$ is unique for the homogeneous and isotropic
universe. Only the solution, Eqs.(\ref{e3}), (\ref{e4}), and (\ref{e5}),
exists as the Machian in the Brans-Dicke theory. In this model, fundamental
quantities of the universe are determined in the framework of the theory.
For the closed model, the value of the relation $GM/c^{2}a=const$ is fixed
to $\pi $. The sign of the gravitational constant $G$\ depends on the
topology of the universe. The gravitational force becomes attractive only
for the closed space, which is hence forgiven physically.\ The universe
expands linearly and all co-moving observers do not accelerate to each other
in this cosmological model.

As for the flat space, another type of solution, Eqs.(\ref{eq1}), (\ref{eq2}%
), and (\ref{e22}) exists for all arbitrary values ($\omega \neq -3/2$) of
the coupling parameter $\omega $. If we request that solutions should exist
for arbitrary values of the coupling parameter (, even though the region is
in general restricted), we may regard that the Machian solution $\phi
=O(\rho /\omega )$ is uniquely determined for the closed, the flat, and the
open spaces respectively in the Brans-Dicke theory.

The scalar field $\phi $\ converges to zero as the universe expands linearly
in the closed model. Taking the conservation law $\rho a^{3}=const$, this is
the unavoidable result in the Machian Point of view. The universe gradually
goes to empty as the universe expands for ever, and particles lose their
inertial properties in otherwise empty universe. The behavior of the scalar
field $\phi (t)\propto t^{-1}$\ means that the gravitational ''constant''
increases linearly ($G(t)\propto t$)\ as the universe expands. However, this
time-variation of the gravitational ''constant'' obviously contradicts with
the results obtained by recent measurements to detect it \cite{8)}\ ($\left| 
\dot{G}/G\right| \lesssim 1.6\times 10^{-12}\,yr^{-1}$).

Moreover, it may be a crucial defect that the coupling parameter $\omega $
must be negative ($\omega <-2$) in this closed model. According to particle
physics, this means that the scalar field becomes the ghost, of which the
energy-momentum is negative. This result is also unavoidable in the Machian
point of view. The discussions for the case of $k=+1$ and $\omega <-2$ are
parallel to those of\ $k=-1$ and $\omega >-2$. The two cases split the sign
of the gravitational constant. The gravitation becomes an attractive force
only for $\omega <-2$ in this Machian solution for the homogeneous and
isotropic universe in the Brans-Dicke theory.\newline
\newline
\textbf{Acknowledgment}

The author is grateful to Professor Carl Brans for helpful discussions and
his hospitality at Loyola University (New Orleans) where this work was done.
He would also like to thank the Nagasaki Prefectural Government for
financial support.

\end{document}